\documentclass[12pt]{article}
\input epsf.sty
\def\mref#1{(\ref{#1})}
\def\eqref#1{(\ref{#1})}
\usepackage{amsfonts}
\usepackage{amssymb}
\newcommand{\tH}{\theta}

\newcommand{\Di}{\Delta_m}
\newcommand{\Dj}{\Delta_n} 
\newcommand{\Dim}{\Delta_{-m}}
\newcommand{\Djm}{\Delta_{-n}}

\newcommand{\vt}{\vartheta}
\newcommand{\G}{\Gamma}
\begin{document}

%%%%%%%%%%%%%%%%%%%%%%%%%%%%%%%%%%%%%%%%%%%%%%%%%%%%%%%%%%%%%%%%%%%
%%% Title Etc.
%%%%%%%%%%%%%%%%%%%%%%%%%%%%%%%%%%%%%%%%%%%%%%%%%%%%%%%%%%%%%%%%%%%

\title{Darboux 
%type 
transformations for  5-point and 7-point self-adjoint schemes and 
an integrable discretization of the 2D Schr\"odinger operator
}
\author{
M. Nieszporski\thanks{Instytut Fizyki Teoretycznej,
Uniwersytet w Bia{\l}ymstoku,
ul. Lipowa 41, 15-424 Bia{\l}ystok, Poland
 e-mails: maciejun@fuw.edu.pl, maciejun@alpha.uwb.edu.pl
tel: 48-85-7457239, fax: 48-85-7457238}, 
P.M. Santini\thanks{Dipartimento di Fisica, Universit\`a di Roma ``La Sapienza'' and 
Istituto Nazionale di Fisica Nucleare, Sezione di Roma, 
Piazz.le Aldo Moro 2, I--00185 Roma, Italy
e-mail: paolo.santini@roma1.infn.it}~ and
A. Doliwa\thanks{Uniwersytet Warminsko-Mazurski w Olsztynie
Wydzial Matematyki i Informatyki
ul. Zolnierska 14 A, 10-561 Olsztyn, Poland
e-mail: doliwa@matman.uwm.edu.pl}
}

\maketitle

\begin{abstract}
With this paper we begin an investigation of difference schemes
that possess Darboux transformations and can be regarded as natural 
discretizations of elliptic partial differential equations.  
We construct, in particular, the Darboux transformations for the general 
self adjoint schemes with five and seven neighbouring points. 
We also introduce a distinguished discretization of the two-dimensional 
stationary 
Schr\"odinger equation, described by a 5-point difference scheme involving two 
potentials, which admits a Darboux transformation.
\end{abstract} 


%%%%%%%%%%%%%%%%%%%%%%%%%%%%%%%%%%%%%%%%%%%%%%%%%%
\section{Introduction}
%%%%%%%%%%%%%%%%%%%%%%%%%%%%%%%%%%%%%%%%%%%%%%%%%%

Linear differential operators admitting Darboux Transformations (DTs) play 
a crucial role in the theory of integrable systems \cite{AS,MNPZ}. i) One 
can associate with 
such operators integrable nonlinear partial differential equations (PDEs). 
ii) One can make use of the 
spectral theory of such linear operators to solve classical initial - 
boundary value problem for the nonlinear PDEs. iii) One can construct 
solutions of the nonlinear equations from simpler solutions, using the 
Darboux - B\"acklund transformations. iv) One can often associate with the 
nonlinear PDE a geometric meaning, inherited by the geometric properties of 
the linear operator.

It is natural to search for a discretization of this beautiful picture.  
In general, searching for distinguished discretizations of linear 
differential operators that admit general DTs (integrable discretizations) 
is not a 
trivial task. Several methods of "integrable discretization" 
have been used so far (see e.g. \cite{Suris}), but no one is fully 
satisfactory. Surprisingly enough, a distinguished integrable 
discretization of the  
two-dimensional stationary Schr\"odinger equation is, to the best of our 
knowledge,  not present in the literature and one of our goals is to change 
this situation.

In recent years the study of linear difference equations that admit
DTs were undertaken. 
Most of the results are based on the 4-point difference scheme, i.e. a 
scheme that relates four neighbouring points
\begin{equation}
\label{4point}
\psi_{m+1,n+1}=\alpha_{m,n} \psi_{m+1,n} + \beta_{m,n} \psi_{m,n+1} +
\gamma_{m,n} \psi_{m,n}
\end{equation}
(where  $\alpha,~\beta,~\gamma$ are real functions of two discrete variables 
$(m,n)\in {\mathbb Z}^2$ and where the standard 
notation $f_{m,n}=f(m,n)$ is used throughout the paper),
which is proper for discretizing 
second order hyperbolic equations in the canonical form. 
In particular, the proper discrete analogue of the Laplace equation for  
conjugate nets
\begin{equation}
\label{Conj}
\Psi,_{uv}+ C \Psi,_{u}+D \Psi,_{v}= 0,
\end{equation}
whose DTs were obtained in \cite{Jonas,Eisenhart}, turns out to be 
the 4-point scheme \cite{BK} 
\begin{equation}
\label{QL}
\psi_{m+1,n+1}=\alpha_{m,n} \psi_{m+1,n} + \beta_{m,n} \psi_{m,n+1} +
(1-\alpha_{m,n}-\beta_{m,n}) \psi_{m,n},
\end{equation}
describing a lattice with planar quadrilaterals \cite{Doliwa}, \cite{DS}, 
whose general DTs 
were extensively studied in recent years (see, e.g., \cite{MDS,KS,DSM}). 
While the Moutard equation \cite{Moutard}
\begin{equation}
\label{Mout}
\Psi,_{uv}=F \Psi,
\end{equation}
relevant in the description of asymptotic nets, 
splits naturally in the discrete case into two equations \cite{NimSch,N}
\begin{equation}
\label{dMout}
\begin{array}{l}
\psi_{m+1,n+1}+\psi_{m,n}= f_{m,n}( \psi_{m+1,n} +  \psi_{m,n+1})\\ 
\psi_{m+1,n+1}+\psi_{m,n}= f_{m+1,n} \psi_{m+1,n} + f_{m,n+1} \psi_{m,n+1}
\end{array}
\end{equation}
admitting DTs. Laplace transformations for the general 4-point scheme (\ref{4point}) 
are also known \cite{Doliwa,Novikov2,Adler}.

The discretization of elliptic equations in two dimensions admitting 
general DTs is much less studied and understood. 
In the continuous case, elliptic equations and their Darboux transformations   
are often obtained from the hyperbolic ones regarding $(u,v)$ as complex 
variables:
\begin{equation}
\label{complex}
u=x+iy \qquad v=x-iy, 
\end{equation}
but this approach presents some problems in the discrete case. 

In our opinion, 
a proper discretization of an elliptic operator should satisfy two 
basic properties. 
\begin{itemize}
\item It should be applicable to solve generic Dirichlet 
boundary value problems on a 2D lattice; 
\item It should possess 
a class of DTs which must be (at least) as rich as that of its differential  
counterpart.
\end{itemize} 

It is very easy to convince one-self that the first criterion cannot be 
satisfied by the 4-point scheme (\ref{4point}) and 
that at least 5-point difference schemes should be 
introduced \cite{Hil}. The most general 5-point scheme reads as follows:
\begin{equation}
\begin{array}{l}
\label{Lpsi}
a_{m,n} \psi_{m+1,n} +(a_{m-1,n}+w_{m,n}) \psi_{m-1,n}
+b_{m,n} \psi_{m,n+1} +(b_{m,n-1}+z_{m,n})  \psi_{m,n-1}=\\
f_{m,n} \psi_{m,n}, 
\end{array} 
\end{equation}
where  $a,~b,~w,~z$ are functions of the discrete variables 
$(m,n)\in {\mathbb Z}^2$. 

\begin{center}
\mbox{\hsize6cm \vbox{\epsfxsize=6cm 
\epsffile{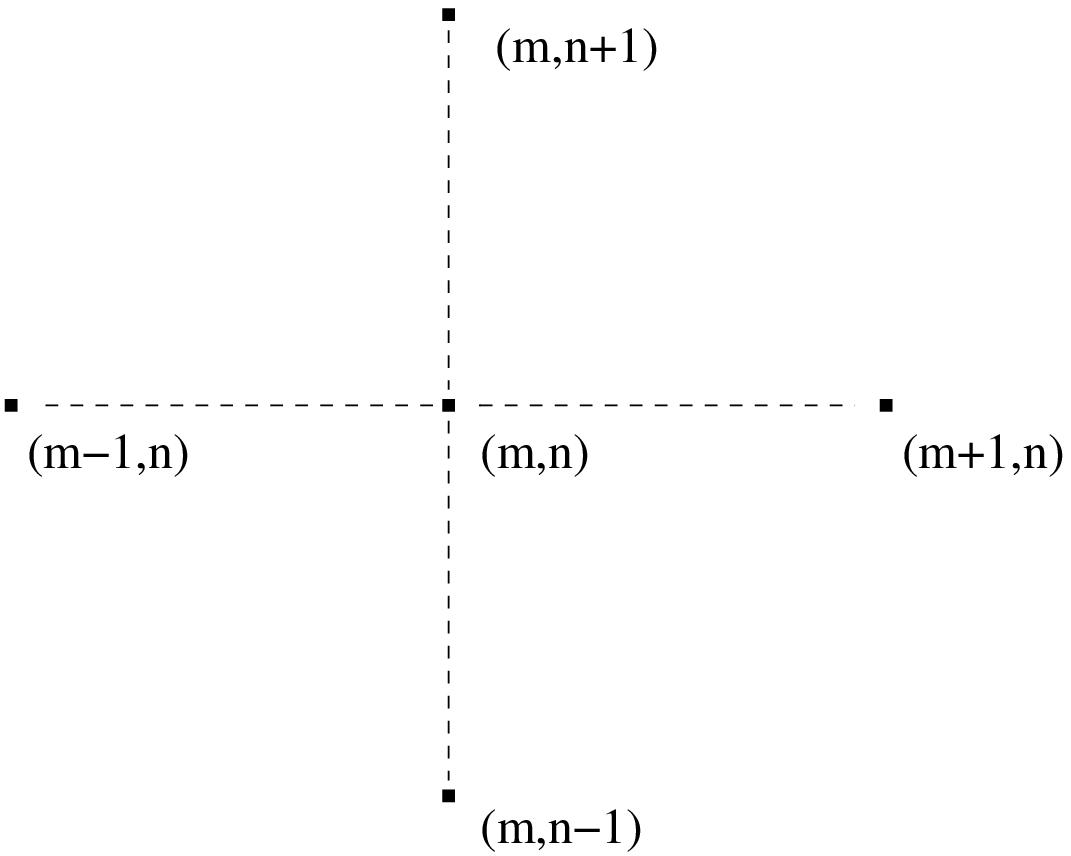}
Fig.1 $~~~~$ The 5 - point stencil}
}
\end{center}
\vskip 10pt
\noindent
Its natural continuous limit
\begin{equation}
\label{limit}
(\epsilon m,\epsilon n)~\to~(x,y),
\end{equation}
in which the lattice spacing $\epsilon$ goes to zero and
\begin{equation}
\label{limit1}
\left(
\begin{array}{c}
\psi_{m,n} \\
a_{m,n}    \\
b_{m,n}    
\end{array}
\right)
=
\left(
\begin{array}{c}
\Psi(x,y) \\
A(x,y)    \\
B(x,y)    
\end{array}
\right) + O(\epsilon^3),~
\left(
\begin{array}{c}
w_{m,n} \\
z_{m,n}     
\end{array}
\right)
=
-\epsilon\left(
\begin{array}{c}
W(x,y) \\
Z(x,y)       
\end{array}
\right) + O(\epsilon^3),
\end{equation}
$$f_{m,n}=2(A+B)-\epsilon (A,_{x}+B,_{y}+W+Z)+
\epsilon^2[F(x,y)+\frac{1}{2}(A,_{xx}+B,_{yy})]+O(\epsilon^3), $$
gives the following second order PDE in the independent variables $(x,y)$
\begin{equation}
\label{Lpsic}
(A \Psi,_x),_x+W \Psi,_x+ (B \Psi,_y),_y +Z \Psi,_y=F\Psi,
\end{equation}
in which the $\Psi,_{xy}$ term is missing (we remark that the $\Psi,_{xy}$ 
term is instead the only second order term present in equations (\ref{Conj}) 
and (\ref{Mout})). Equation (\ref{Lpsic}) is elliptic if $AB>0$ and its 
canonical form is obtained by setting $A=1=B$. The functions $w,z,W,Z$ 
measure the departure of equations (\ref{Lpsi}) and (\ref{Lpsic}) from 
self-adjointness, which is a basic property in most of the applications. 
  
In this paper we present the following results.
\begin{itemize}
\item We construct the DTs 
for the self-adjoint reductions of the 5-point scheme  \mref{Lpsi} and of  
its continuous limit \mref{Lpsic}. 
\item We use the gauge covariance property of the general self-adjoint 
5-point scheme to construct a distinguished integrable discrete analogue of 
the stationary Schr\"odinger 
operator in two dimensions. ``Integrable'' in the sense that, not only 
it reduces to the stationary Schr\"odinger operator in the continuous limit, 
but also it possesses DTs. 
\item We show that the construction of DTs for the self-adjoint 5-point 
scheme can be applied to the case of self-adjoint schemes involving 
more neighbouring points, illustrating such a generalization in the case 
of the general self-adjoint 7-point scheme, for which Laplace 
transformations are already known \cite{Novikov,Novikov2}.    
\end{itemize}

The paper is organized as follows. Section \ref{SAD} is devoted to the 
construction of the DTs for the most general 
self-adjoint 5-point difference operator with complex coefficients:
\begin{equation}
\label{selfa}
{\mathcal L}_{5}=a_{m,n}T_m+a_{m-1,n}T^{-1}_m+b_{m,n}T_n+b_{m,n-1}T^{-1}_n-
f_{m,n},
\end{equation}
where $T_m$ and $T_n$ are the translation operators with respect to the 
discrete variables $(m,n)\in{\mathbb Z}^2$:
\begin{displaymath}
T_mf_{m,n}=f_{m+1,n},~~~T_nf_{m,n}=f_{m,n+1}
\end{displaymath}
and $a,~b,~c$ are complex functions of the two discrete variables 
$(m,n)\in {\mathbb Z}^2$.

We also use the following difference operators
\begin{displaymath}
\begin{array}{l}
\Di f_{m,n}=f_{m+1,n}-f_{m,n} \qquad \Dj f_{m,n}=f_{m,n+1}-f_{m,n} \\
 \Dim f_{m,n}=f_{m-1,n}-f_{m,n} \qquad \Djm f_{m,n}=f_{m,n-1}-f_{m,n}.
\end{array}
\end{displaymath}
The operator ${\mathcal L}_{5}$ is formally 
self-adjoint with respect to the bilinear form:
\begin{equation}
<f,g>:=\sum_{m,n}f_{m,n}g_{m,n}
\end{equation}
and, in the particular case $a=b=1$, it reduces to the 5-point scheme 
\begin{equation}
\label{ds}
{\mathcal L}_{Sch}=T_m+T^{-1}_m+T_n+T^{-1}_n-f,
\end{equation}
which is the simplest and most used  (in numerical applications) 
discretization of the Schr\"odinger operator 
\begin{equation}
\label{schr}
{\partial^2\over \partial x^2}+{\partial^2\over \partial y^2}-F,
\end{equation}
while the further constraint
$f=4$ leads to 
\begin{equation}
\label{dl}
{\mathcal L}_0=T_m+T^{-1}_m+T_n+T^{-1}_n-4,
\end{equation}
i.e. to
a discretization of the Laplacian  
${\partial^2\over \partial x^2}+{\partial^2\over \partial y^2}$ 
often used in the numerical  studies  of elliptic boundary value problems 
\cite{Hil}.

In the continuous limit, the operator ${\mathcal L}_{5}$ goes 
to the second order partial differential operator 
\begin{equation}
\label{selfA}
\begin{array}{l}
L=A\frac{\partial^2}{\partial x^2}+B\frac{\partial^2}{\partial y^2}+
A,_x\frac{\partial}{\partial x}+B,_y \frac{\partial}{\partial y} -F
\end{array}
\end{equation}
which  is formally self-adjoint with respect to the bilinear form:
\begin{equation}
<f,g>:=\int fg dx dy.
\end{equation}

We notice that, while both operators (\ref{selfa}), (\ref{selfA}) are   
self-adjoint, the two ``integrable'' discretizations
\mref{dMout} of the self-adjoint Moutard equation \mref{Mout} are just 
mutually adjoint.

In section \ref{SCH}, using the following covariance property 
(gauge invariance) of the operator ${\mathcal L}_{5}$: 
\begin{equation}
\begin{array}{c}
\label{gaugea}
{\mathcal L}_{5} \to \tilde{\mathcal L}_{5} =  
g_{m,n} {\mathcal L}_{5} g_{m,n} \\
a_{m,n} \to \tilde{a}_{m,n} = a_{m,n} g_{m,n} g_{m+1,n}, \qquad
b_{m,n} \to \tilde{b}_{m,n} = b_{m,n} g_{m,n} g_{m,n+1}, \\
f_{m,n} \to \tilde{f}_{m,n} = f_{m,n} g^2_{m,n},
\end{array}
\end{equation}
we draw the operator  ${\mathcal L}_{5}  $ to the form
\begin{equation}
\label{schr5}
{\mathcal L}_{SchInt}=\frac{\G_{m,n}}{\G_{m+1,n}} T_m+
\frac{\G_{m-1,n}}{\G_{m,n}}T^{-1}_m+
\frac{\G_{m,n}}{\G_{m,n+1}} T_n+\frac{\G_{m,n-1}}{\G_{m,n}}T^{-1}_n-f.
\end{equation}
The operator \mref{schr5}, symmetric with respect to $\pi /2$ rotations 
in the plane grid $(m,n)$, reduces to the two-dimensional Schr\"odinger 
operator 
\mref{schr} under the natural continuous limit and possesses DTs. Therefore it 
appears as the ``integrable'' discrete analogue of the two-dimensional 
Schr\"odinger operator. It is therefore the proper starting  point in the 
search for the integrable discrete analogues of the nonlinear PDEs 
of elliptic type associated with \mref{schr} (like the Veselov-Novikov 
hierarchy \cite{VN}, nonlinear $\sigma$ models \cite{Misner} and the Ernst equation \cite{Ernst}).  

The corresponding gauge transformation for the operator $L$ in (\ref{selfA}) 
is 
\begin{equation}
\label{Gs}
\begin{array}{c}
L \rightarrow \tilde{L}= g L g, \\
\tilde{A}=g^2A, \qquad \tilde{B}=g^2B, \\
\tilde{F}=gLg.
\end{array}
\end{equation}

We draw the reader's attention to the fact that constraint $A=B$ is conserved 
by the gauge transformation
\mref{Gs} and the same is true for the constraint 
$a_{m,n+1}a_{m,n} = b_{m+1,n} b_{m,n}$ with respect to gauge
\mref{gaugea}. In the case of the reduction $A=B$, the operator \mref{selfA} 
can be gauged into the form of the 
2D Schr\"odinger operator \mref{schr} while, in the case $a_{m,n+1}a_{m,n} = 
b_{m+1,n} b_{m,n}$,  
the operator \mref{selfa} can be gauged into the operator \mref{ds}. But 
while there exist DTs
which preserves the form of equation \mref{schr}, we don't know 
DTs that preserves the form of equation
 \mref{ds}. This is the reason why we view the operator \mref{schr5} and not 
the operator \mref{ds} as the integrable discretization of the Schr\"odinger 
equation \mref{schr}.

Finally, in section \ref{SAD7} we show that there is another self-adjoint scheme
involving more neighbouring points which possess DTs of the type presented in 
section \ref{SAD}. It is a  self-adjoint 
7-point difference scheme

\begin{equation}
\label{selfa7}
\begin{array}{l}
{\mathcal L}_{7}\psi_{m,n}:=
a_{m,n}\psi_{m+1,n}+a_{m-1,n}\psi_{m-1,n}+b_{m,n}\psi_{m,n+1}+ \\
b_{m,n-1}\psi_{m,n-1}+ 
s_{m+1,n} \psi_{m+1,n-1}+ 
s_{m,n+1} \psi_{m-1,n+1}-f_{m,n}\psi_{m,n}=0,
\end{array}
\end{equation}
which, in the continuous limit 
\begin{equation}
\label{limit2}
\left(
\begin{array}{c}
\psi_{m,n} \\
a_{m,n}+s_{m,n}    \\
b_{m,n}+s_{m,n}    \\
s_{m,n}    
\end{array}
\right)
=
\left(
\begin{array}{c}
\Psi(x,y) \\
A(x,y)    \\
B(x,y)    \\
-S(x,y) 
\end{array}
\right) + O(\epsilon^3),
\end{equation}
$$f_{m,n}=2(A+B+S)-\epsilon (A,_{x}+B,_{y}+2S,_{x}+2S,_{y})+
\epsilon^2[F(x,y)+\frac{1}{2}(A,_{xx}+B,_{yy})]+O(\epsilon^3) $$
goes to the most general second order self-adjoint differential equation in 
two independent variables:
\begin{equation}
\label{selfa7c}
(A \Psi,_x),_x+(S \Psi,_y),_x+ (B \Psi,_y),_y +(S \Psi,_x),_y=F\Psi.
\end{equation}

The operator ${\mathcal L}_{7}$ in (\ref{selfa7}), introduced in 
\cite{Novikov}, admits the representation ${\mathcal L}_{7}=QQ^++w$ in 
terms of a 3-point difference operator $Q$ and of its adjoint $Q^+$, and 
this factorization plays a crucial role in the construction of Laplace 
transformations for ${\mathcal L}_{7}$ \cite{Novikov,Novikov2} and in the 
development of an associated discrete complex function theory \cite{Novikov3}. 
The 5-point operator ${\mathcal L}_{5}$ does not admit the above 
representation and the construction of the DTs of this paper does not make 
use of it.   
We also remark that several discretizations of the 1D Schr\"odinger 
operator have appeared in the literature throughout the years 
(see, f.i., \cite{flaschka,Manakov,Shabat} and \cite{Novikov2} with references therein included).   
We end this introduction observing that a 4-point scheme 
with a complexification of the discrete variables $(m,n)$ analogous to 
\mref{complex}, was used in \cite{GLM} to construct a discrete analogue of 
the $\sigma$ model.  

%%%%%%%%%%%%%%%%%%%%%%
%%%%%%%%%%%%%%%%%%%%%%
\section{5-point self adjoint  operator and its Darboux transformation}
%%%%%%%%%%%%%%%%%%%%%%
%%%%%%%%%%%%%%%%%%%%%%
\label{SAD}
In this section we present a Darboux transformation for the equation:

\begin{equation}
\label{saeq}
a_{m,n} \psi_{m+1,n} +a_{m-1,n} \psi_{m-1,n}
+b_{m,n} \psi_{m,n+1} +b_{m,n-1}  \psi_{m,n-1}=f_{m,n} \psi_{m,n},
\end{equation}
%\begin{equation}
%\label{saeq}
%\U a_{ \, \U x + a \, \D x + \R b \, \R x+ b \, \Le x =F \, x 
%\end{equation}
where  $a_{m,n}$,  $b_{m,n}$ and $f_{m,n}$ are  given functions.
Let  $\tH$ be another solution of (\ref{saeq}), i.e:
\begin{equation}
\label{normal1}
a_{m,n} \tH_{m+1,n} +a_{m-1,n} \tH_{m-1,n}
+b_{m,n} \tH_{m,n+1} +b_{m,n-1}  \tH_{m,n-1}=f_{m,n} \tH_{m,n};
\end{equation}
then:
\begin{equation}
\label{ff}
f_{m,n}=\frac{1}{\tH}\left(a_{m,n} \tH_{m+1,n} +a_{m-1,n} \tH_{m-1,n}
+b_{m,n} \tH_{m,n+1} +b_{m,n-1}  \tH_{m,n-1}\right).
\end{equation}
%\begin{equation}
%\label{ssaeq}
%\U a \, \U \tH + a \, \D \tH + \R b \, \R \tH + b \, \Le \tH =F \,
%\tH. 
%\end{equation}
Eliminating $f_{m,n}$ from \mref{saeq} and  \mref{normal1} we get
\begin{equation}
%\label
\begin{array}{l}
\Di (a_{m-1,n} \psi_{m,n}  \tH_{m-1,n} -a_{m-1,n} \tH_{m,n}  \psi_{m-1,n})+\\
\Dj (b_{m,n-1} \psi_{m,n}  \tH_{m,n-1}- b_{m,n-1} \tH_{m,n}  \psi_{m,n-1})=0.
\end{array}
\end{equation}
It means that there exists  a function $\alpha$ such that
\begin{equation}
\begin{array}{l}
%\label{}
\Dj \alpha = a_{m-1,n} \tH_{m,n} \tH_{m-1,n} 
\Dim \frac{\psi_{m,n}}{\tH_{m,n}},
\\
\Di \alpha = -b_{m,n-1} \tH_{m,n}  \tH_{m,n-1} 
\Djm \frac{\psi_{m,n}}{\tH_{m,n}}.
\end{array}
\end{equation}
Setting \[\psi'_{m,n}=\frac{\alpha_{m,n}}{\tH_{m,n}}\]
we find that $\psi'_{m,n}$ satisfies the following equation
\begin{equation}
\label{saeqp}
 a'_{m,n} \,  \psi'_{m+1,n} + a'_{m-1,n} \,  
\psi'_{m-1,n} +  b'_{m,n} \,  \psi'_{m,n+1}+ b'_{m,n-1} \, 
 \psi'_{m,n-1} =f'_{m,n} \, \psi'_{m,n}, 
\end{equation}
where
\begin{equation}
%\label{}
a'_{m-1,n}=\frac{\tH_{m,n}}{ b_{m-1,n-1} \,   \tH_{m-1,n-1}} \qquad
b'_{m,n-1}=\frac{\tH_{m,n}}{ a_{m-1,n-1} \,  \tH_{m-1,n-1}}
\end{equation}
and
\begin{equation}
\label{ff'}
f'_{m,n}=\tH_{m,n} \left( 
 a'_{m,n} \,  \frac{1}{\tH_{m+1,n}} + a'_{m-1,n} \,  \frac{1}{\tH_{m-1,n}} +
  b'_{m,n} \,  \frac{1}{\tH_{m,n+1}}+ b' _{m,n-1} \,  
\frac{1}{\tH_{m,n-1}}\right).
\end{equation}
Comparing equations (\ref{ff}) and (\ref{ff'}), we also infer that 
$\theta'=1/\theta$ is a solution of (\ref{saeqp}).
 
In the continuous limit (\ref{limit}), with 
$\theta_{m,n}=\Theta (x,y)+O(\epsilon^3)$ and with $f_{m,n}$ expanded 
according to (\ref{ff}) to get (\ref{limit1}b), we obtain the DT 
\begin{equation}
%\label{}
\frac{1}{\Theta} (\Theta \Psi'),_y =-A \Theta (\frac{\Psi}{\Theta}),_x
\qquad
\frac{1}{\Theta} (\Theta \Psi'),_x =B \Theta  (\frac{\Psi}{\Theta}),_y
\end{equation}
from the solution space of equation
\begin{equation}
\label{sacl}
(A \Psi,,_x),_x + (B \Psi,_y),_y =F\Psi
\end{equation}
to the solution space of equation
\begin{equation}
\label{sacl'}
(A' \Psi',_x),_x + (B' \Psi',_y),_y =F'\Psi',
\end{equation}
where $\Theta$ is another solution of \mref{sacl}, so that
\begin{equation}
\label{F}
F= \frac{1}{\Theta}[(A \Theta,_x),_x + (B \Theta,_y),_y ],
\end{equation}
and the new "potentials" $A'$, $B'$ and $F'$ are related to the old ones 
as follows
\begin{equation}
\label{A'}
\begin{array}{c}
A'=\frac{1}{B}, \qquad B'=\frac{1}{A}, \\
F'=\Theta [ (A' (\frac{1}{\Theta}),_x),_x + (B' (\frac{1}{\Theta}),_y),_y].
\end{array}
\end{equation}
Comparing equations (\ref{sacl'}) and (\ref{A'}c), we also infer that 
$\Theta'=1/\Theta$ is a solution of (\ref{sacl'}).          
%%%%%%%%%%%%%%%%%%%%%%
%%%%%%%%%%%%%%%%%%%%%%
\section{A two-dimensional Schr\"odinger operator and its Darboux 
transformation}
%%%%%%%%%%%%%%%%%%%%%%
%%%%%%%%%%%%%%%%%%%%%%
\label{SCH}
The operator \mref{selfa} can be gauged into the form \mref{schr5}. Indeed,
if $a_{m,n}$ and $b_{m,n}$ are given functions, one can find a function 
$g_{m,n}$ such that $\frac{a_{m,n}}{b_{m,n}}=\frac{g_{m,n+1}}{g_{m+1,n}}$ and, 
under the gauge \mref{gaugea}, we get 
$\tilde{a}_{m,n}=\tilde{b}_{m,n}=:\frac{1}{\Gamma^2_{m,n}}$. Finally, the 
operator $\hat{\mathcal L}$ given by
$ \hat{\mathcal L}_{5} =  
\frac{1}{\Gamma_{m,n}}\tilde{{\mathcal L}}_{5} \frac{1}{\Gamma_{m,n}}$
is of the wanted form \mref{schr5}.

Combining the DT of the previous section with the above gauge transformation, 
one obtains the following DT for the discrete analogue \mref{schr5} of the 
Schr\"odinger operator.

Let $N_{m,n}$ be a solution of the integrable discrete analogue of the 
2D Schr\"odinger operator:
\begin{equation}
\label{Seq}
\begin{array}{l}
\frac{\G_{m,n}}{\G_{m+1,n}} N_{m+1,n} +
\frac{\G_{m-1,n}}{\G_{m,n}} N_{m-1,n} +
\frac{\G_{m,n}}{\G_{m,n+1}} N_{m,n+1} +
\frac{\G_{m,n-1}}{\G_{m,n}} N_{m,n-1} 
=f_{m,n} N_{m,n},
\end{array}
\end{equation}
and let $\vt$ be another solution of it. It means that
\begin{equation}
\label{f}
f_{m,n} =\frac{1}{\vt}\left(\frac{\G_{m,n}}{\G_{m+1,n}} \vt_{m+1,n} +
\frac{\G_{m-1,n}}{\G_{m,n}} \vt_{m-1,n} +
\frac{\G_{m,n}}{\G_{m,n+1}} \vt_{m,n+1} +
\frac{\G_{m,n-1}}{\G_{m,n}} \vt_{m,n-1}\right).
\end{equation}
Eliminating $f_{m,n}$ from \mref{f} and \mref{Seq}
we get that there exists a function $N'_{m,n}$, given in quadratures by
\begin{equation}
\label{Mt}
\begin{array}{l}
\Di \left( \frac{\vt_{m,n}}{\G_{m,n}} \G'_{m,n} N'_{m,n} \right)=
\frac{\G_{m,n-1}}{\G_{m,n}}  \vt_{m,n} \vt_{m,n-1} 
\left(\frac{N_{m,n}}{\vt_{m,n}} -\frac{N_{m,n-1}}{\vt_{m,n-1}} \right) 
%\Djt \frac{N_{m,n}}{\vt_{m,n}},
\\
\Dj \left( \frac{\vt_{m,n}}{\G_{m,n}} \G'_{m,n} N'_{m,n} \right)=
-\frac{\G_{m-1,n}}{\G_{m,n}}  \vt_{m,n} \vt_{m-1,n}  
\left(\frac{N_{m,n}}{\vt_{m,n}} -\frac{N_{m-1,n}}{\vt_{m-1,n}} \right)
%\Dit \frac{N_{m,n}}{\vt_{m,n}},
\end{array}
\end{equation}
%where
%\begin{equation}
%\Dit:=1-(T_1)^{-1} \qquad \Djt:=1-(T_2)^{-1}.
%\end{equation}
which satisfies
\begin{equation}
\label{Seqp}
\begin{array}{l}
\frac{\G'_{m,n}}{\G'_{m+1,n}} N'_{m+1,n} +
\frac{\G'_{m-1,n}}{\G'_{m,n}} N'_{m-1,n} +
\frac{\G'_{m,n}}{\G'_{m,n+1}} N'_{m,n+1} +
\frac{\G'_{m,n-1}}{\G'_{m,n}} N'_{m,n-1} 
=f'_{m,n} N'_{m,n},
\end{array}
\end{equation}
where
\begin{equation}
\label{Mtc}
\begin{array}{l}
\G'^2_{m,n}=\frac{\G_{m,n}\G_{m-1,n-1} \vt_{m-1,n-1}}{\vt_{m,n}}, 
%\frac{\sqrt{\G_{m-1,n-1} \vt_{m-1,n-1}}}{\sqrt{\G_{m,n} \vt_{m,n}}},
\\
f'_{m,n}=\frac{\G' \vt}{\G} 
\left(
\frac{\G'_{m,n}}{\G'_{m+1,n}} (\frac{\G}{\G' \vt})_{m+1,n} +
\frac{\G'_{m-1,n}}{\G'_{m,n}} (\frac{\G}{\G' \vt})_{m-1,n} +
\frac{\G'_{m,n}}{\G'_{m,n+1}} 	(\frac{ \G}{\G'\vt})_{m,n+1} +\right. \\
\left.\frac{\G'_{m,n-1}}{\G'_{m,n}} (\frac{ \G}{\G'\vt})_{m,n-1}
\right) = 
(\G\vt)_{m-1,n-1}[
\frac{1}{(\G\vt)_{m-1,n}}+\frac{1}{(\G\vt)_{m,n-1}}]
+\frac{\vt}{\G}[(
\frac{\G}{\vt})_{m-1,n}+(
\frac{\G}{\vt})_{m,n-1}] .
\end{array}
\end{equation}
Now the function $\vt'=\frac{\G}{\G'\vt}$ is a solution of \mref{Seqp}.
In the continuous limit, with $\Gamma_{m,n}=G(x,y)+O(\epsilon)$ 
and with $f_{m,n}$ expanded according to equation (\ref{f}), one 
obtains the classical Moutard transformation
\begin{equation}
\frac{1}{\Theta}(\Theta N'),_x=\Theta\left(\frac{N}{\Theta}\right),_y~,~~~~
\frac{1}{\Theta}(\Theta N'),_y=-\Theta\left(\frac{N}{\Theta}\right),_x
\end{equation} 
from the solution space of equation
\begin{equation}
\label{aa}
\Psi,_{xx}+\Psi,_{yy}=F\Psi
\end{equation}
to the solution space of equation 
\begin{equation}
\label{aap}
\Psi',_{xx}+\Psi',_{yy}=F'\Psi',
\end{equation}
where $\Theta$ is another solution of \mref{aa}, so that 
$$F=\frac{1}{\Theta}(\Theta,_{xx}+\Theta,_{yy}),$$
and 
$$F'=\Theta [\left(\frac{1}{\Theta}\right),_{xx}+
\left(\frac{1}{\Theta}\right),_{yy}].$$
Furthermore $\Theta'=1/\Theta$ is a solution of \mref{aap}.

%%%%%%%%%%%%%%%%%%%%%%
%%%%%%%%%%%%%%%%%%%%%%
\section{7-point self-adjoint operator and its Darboux 
%type 
transformations}
%%%%%%%%%%%%%%%%%%%%%%
%%%%%%%%%%%%%%%%%%%%%%
\label{SAD7}
We end this paper showing that the construction of DTs presented in 
the previous two sections applies also to a self-adjoint scheme involving 
more than 5 points, namely  in the case of the  
self-adjoint 7-point scheme:
\begin{equation}
\label{sa7}
\begin{array}{l}
a_{m,n} \psi_{m+1,n} +a_{m-1,n} \psi_{m-1,n}
+b_{m,n} \psi_{m,n+1} +b_{m,n-1}  \psi_{m,n-1}+\\
s_{m+1,n} \psi_{m+1,n-1}+s_{m,n+1} \psi_{m-1,n+1}=
f_{m,n} \psi_{m,n},
\end{array}
\end{equation}
where
$a_{m,n}$,  $b_{m,n}$, $s_{m,n}$ and $f_{m,n}$ are given functions, 
%(the reality of these function is necessary for the self-adjointness, but is 
%irrelevant in the construction of the DT)
which, as we have seen in the 
introduction, is a discretization of the most general second order, 
self-adjoint, linear, differential equation in two independent variables.

Let  $\tH_{m,n}$ be another solution of equation \mref{sa7}:
\begin{equation}
\label{sas7}
\begin{array}{l}
a_{m,n} \tH_{m+1,n} +a_{m-1,n} \tH_{m-1,n}
+b_{m,n} \tH_{m,n+1} +b_{m,n-1}  \tH_{m,n-1}
+\\
s_{m+1,n} \tH_{m+1,n-1}+s_{m,n+1} \tH_{m-1,n+1}
=f_{m,n} \tH_{m,n}.
\end{array}
\end{equation}
%\begin{equation}
%\label{ssaeq}
%\U a \, \U \tH + a \, \D \tH + \R b \, \R \tH + b \, \Le \tH =F \,
%\tH. 
%\end{equation}
Eliminating $f_{m,n}$ from \mref{sa7} and  \mref{sas7} we get
\begin{equation}
%\label
\begin{array}{l}
\Di [a_{m-1,n}  \tH_{m,n}  \tH_{m-1,n} (\frac{\psi_{m,n}}{\tH_{m,n}} - 
\frac{\psi_{m-1,n}}{\tH_{m-1,n} })
+ s_{m,n} \tH_{m-1,n} \tH_{m,n-1} (\frac{\psi_{m,n-1}}{\tH_{m,n-1}} - 
\frac{\psi_{m-1,n}}{\tH_{m-1,n} })]+
\\
\Dj [b_{m,n-1} \tH_{m,n}   \tH_{m,n-1} (\frac{\psi_{m,n}}{\tH_{m,n}}- 
\frac{\psi_{m,n-1}}{\tH_{m,n-1}})
+s_{m,n} \tH_{m-1,n} \tH_{m,n-1} (\frac{\psi_{m-1,n}}{\tH_{m-1,n}}- 
\frac{\psi_{m,n-1}}{\tH_{m,n-1}})]=0.
\end{array}
\end{equation}
It means that there exists  a function $\alpha$ such that
\begin{equation}
\begin{array}{l}
%\label{}
\Dj \alpha_{m,n} = \left( a_{m-1,n} \tH_{m,n} \tH_{m-1,n} + s_{m,n} 
\tH_{m-1,n} \tH_{m,n-1}   \right) 
\Dim \frac{\psi_{m,n}}{\tH_{m,n}}
- s_{m,n} \tH_{m-1,n} \tH_{m,n-1} 
\Djm \frac{\psi_{m,n}}{\tH_{m,n}}
\\
\Di \alpha_{m,n} = -(b_{m,n-1} \tH_{m,n}  \tH_{m,n-1} +
s_{m,n} \tH_{m-1,n} \tH_{m,n-1} )
\Djm \frac{\psi_{m,n}}{\tH_{m,n}}
+s_{m,n} \tH_{m-1,n} \tH_{m,n-1} 
\Dim \frac{\psi_{m,n}}{\tH_{m,n}}.
\end{array}
\end{equation}
Introducing \[\psi'_{m,n}=\frac{\alpha_{m,n}}{\tH_{m,n}}\]
we find that $\psi'_{m,n}$ satisfies the following equation
\begin{equation}
\begin{array}{l}
\label{bb}
 a'_{m,n} \,  \psi'_{m+1,n} + a'_{m-1,n} \,  \psi'_{m-1,n} +  b'_{m,n} \,  \psi'_{m,n+1}+
 b'_{m,n-1} \, 
 \psi'_{m,n-1} +\\
s'_{m+1,n} \psi'_{m+1,n-1}+s'_{m,n+1} \psi'_{m-1,n+1}
=f'_{m,n} \, \psi'_{m,n}, 
\end{array}
\end{equation}
where the new fields are given by
\begin{equation}
\begin{array}{lll}
%\label{array}
a'_{m,n}&=&\frac{\tH_{m,n} \tH_{m+1,n} a_{m-1,n}}{\tH_{m,n-1}  p_{m,n}},  \\
b'_{m,n}&=&\frac{\tH_{m,n} \tH_{m,n+1} b_{m,n-1}}{\tH_{m-1,n}  p_{m,n}},\\
s'_{m,n}&=&\frac{s_{m-1,n-1} \tH_{m-1,n} \tH_{m,n-1} }
{\tH_{m-1,n-1} p_{m-1,n-1}}, \\
f'_{m,n}&=&\tH_{m,n} (a'_{m,n} \frac{1}{\tH_{m+1,n}}+   
a'_{m-1,n} \frac{1}{\tH_{m-1,n}}+  
b'_{m,n} \frac{1}{\tH_{m,n+1}} + b'_{m,n-1} \frac{1}{\tH_{m,n-1}}+\\
&&s'_{m+1,n} \frac{1}{\tH_{m+1,n-1}}+s'_{m,n+1} \frac{1}{\tH_{m-1,n+1}})
\end{array}
\end{equation}
and where
$p_{m,n}= \tH_{m,n}a_{m-1,n}b_{m,n-1} +
\tH_{m-1,n} s_{m,n} a_{m-1,n}+s_{m,n} \tH_{m,n-1} b_{m,n-1}$. Again 
$\tH'_{m,n}=1/\tH_{m,n}$ is a solution of \mref{bb}.

In the continuous limit (\ref{limit2}), with 
$\tH_{m,n}=\Theta (x,y)+O(\epsilon^3)$ and with $f_{m,n}$ expanded according 
to \mref{sas7} to obtain \mref{limit2}, we obtain the DT 
\begin{equation}
%\label{}
\begin{array}{c}
\frac{1}{\Theta} (\Theta \Psi'),_y =-A \Theta (\frac{\Psi}{\Theta}),_x 
-S \Theta (\frac{\Psi}{\Theta}),_y
\\
\frac{1}{\Theta} (\Theta \Psi'),_x =B \Theta  (\frac{\Psi}{\Theta}),_y
+S \Theta (\frac{\Psi}{\Theta}),_x
\end{array}
\end{equation}
from the solution space of equation
\begin{equation}
\label{saclf}
(A \Psi,_x),_x + (B \Psi,_y),_y +(S \Psi,_y),_x+(S \Psi,_x),_y =F\Psi
\end{equation}
to the solution space of equation
\begin{equation}
\label{cc}
(A' \Psi',_x),_x + (B' \Psi',_y),_y +(S' \Psi',_y),_x+(S' \Psi',_x),_y=F'\Psi',
\end{equation}
where $\Theta$ is fixed solution of \mref{saclf}; so
\begin{equation}
%\label{}
F= \frac{1}{\Theta}[(A \Theta,_x),_x + (B \Theta,_y),_y+(S \Theta,_y),_x+(S \Theta,_x),_y ] 
\end{equation}
and the new "potentials" $A'$, $B'$, $S'$ and $F'$ are related to the old 
ones as follows
\begin{equation}
\begin{array}{c}
A'=\frac{A}{A B-S^2}, \qquad B'=\frac{B}{AB-S^2}, 
\qquad S'=\frac{S}{AB-S^2}, \\
F'=\Theta [ (A' (\frac{1}{\Theta}),_x),_x + (B' (\frac{1}{\Theta}),_y),_y
+(S' (\frac{1}{\Theta}),_x),_y+(S' (\frac{1}{\Theta}),_y),_x ].
\end{array}
\end{equation}
Again $\Theta'=1/\Theta$ is a solution of \mref{cc}. 
\vskip 30pt
\noindent
{\bf Acknowledgments} 
One of us (PMS) acknowledges useful discussions with M.J.Ablowitz.  
This work was 
 supported by the cultural and scientific agreement between 
the University of Roma ``La Sapienza'' and the University of Warsaw and by the 
University of Warmia and Mazury in Olsztyn under the grant  522-1307-0201 and 
partially supported by KBN grant 2 P03B 126 22.


\begin{thebibliography}{c}

\bibitem{AS} 
M.J. Ablowitz and H. Segur, 
{\it Solitons and the Inverse Scattering Transform}, 
SIAM, Phyladelphia, 1981.
 
\bibitem{MNPZ} 
S.V. Manakov, S.P. Novikov, L.P. Pitaevskii and V.E. Zakharov, 
{Theory of Solitons: the Inverse Scattering Method}, 
Consultants Bureau, New York, 1984.

\bibitem{Suris} 
Y.B. Suris,
{\it R-matrices and Integrable Discretization},
in: Discrete integrable geometry and physics, Clarendon Press, Oxford 1999.

\bibitem{Jonas}
H. Jonas, 
{\em \"Uber die Transformation der konjugierten Systeme und \"uber
den gemeinsamen Ursprung der Bianchischen Permutabilit\"atstheoreme},
Berlin Sitzungsber. XIV (1915) 96-118.

\bibitem{Eisenhart}
L.P. Eisenhart, 
{\it Transformation of surfaces},
Princeton University  Press, Princeton 1923.

\bibitem{BK} 
L.V. Bogdanov and B.G. Konopelchenko, 
{\it Lattice and q-difference Darboux-Zakharov-Manakov systems via $\bar\partial$ method}, 
J. Phys A - Math. Gen. 28 (1995) L173-L178.

\bibitem{Doliwa} 
A. Doliwa, 
{\em Geometric discretization of the Toda system}, 
Phys. Lett. A 234 (1997) 187-192.

\bibitem{DS}
A. Doliwa and  P.M. Santini,
   {\it  Multidimensional quadrilateral lattices are integrable},
     Phys. Lett A 233 
%(4-6): 
(1997) 365-372. 

\bibitem{MDS}
M. Manas, A. Doliwa  and P.M. Santini,
{\it     Darboux transformations for multidimensional quadrilateral lattices .1.},
     Phys. Lett. A 232 
%(1-2): 
(1997) 99-105.

\bibitem{KS}
B.G. Konopelchenko and W.K. Schief, 
{\it Three-dimensional integrable lattices in 
Euclidean spaces: Conjugacy and Orthogonality}, Proc. Roy. Soc. London A 454 
(1998) 3075-3104.

\bibitem{DSM}
A. Doliwa, P.M. Santini and M. Manas,
{\it Transformations of quadrilateral lattices}
J. Math. Phys. 41 
%(2): 
(2000) 944-990.


\bibitem{Moutard}
Th-F. Moutard,
{\it Sur la construction des  \'equations de la forme 
$\frac{1}{z} \frac{\partial^2 z}{\partial x\partial y }=\lambda (x,y)$,
qui admettent une integrale g\'en\'eral explicite}
J. Ec. Pol. 45 (1878)~1.

\bibitem{NimSch}
J.J.C. Nimmo and W.K. Schief,
{\it Superposition principles associated with the Moutard transformation: 
an integrable discretization of a 2+1-dimensional sine-Gordon system},
Proc. R. Soc. London A 453 (1997) 255-279.

\bibitem{N}
M. Nieszporski,
{\it   A Laplace ladder of discrete Laplace equations}
Theor. Math. Phys. 133  (2002) 1576-1584.

\bibitem{Novikov2}
S.P. Novikov, I.A. Dynnikov, 
{\it Discrete spectral symmetries of low-dimensional differential operators and 
difference operators on regular lattices and two-dimensional manifolds}, 
Russian Math. Surveys 52 (1997)
%, no. 5, 
1057--1116.

\bibitem{Adler}
 V.E. Adler, S.Ya. Startsev, {\em Discrete analogues of the Liouville equation},
 Theor. Math. Phys. 121 
%(2): 
 (1999) 1484-1495. 

\bibitem{Hil} 
F.B. Hildebrand, 
{\it Finite - difference equations and simulations}, 
Englewood Cliffs, Prentice-Hall (1968).


\bibitem{Novikov} 
S.P. Novikov, 
{\it Algebraic properties of two-dimensional difference operators},
Russian Math. Surveys 52 (1997) 1.

\bibitem{VN} 
A.P. Veselov and S.P. Novikov, 
{\it Finite-zone, two-dimensional potential Schr\"odinger operator. 
Explicit formulas and evolution equations}, 
Soviet Math. Dokl. 30 (1984) 588-591.

\bibitem{Misner}
C.W. Misner,
{\em Harmonic maps as models for physical theories}, 
Phys. Rev. D 18 (1978) 4510-4524.

\bibitem{Ernst} 
F.J. Ernst, 
{\it New Formulation of the Axially Symmetric Gravitational Field Problem}, 
Phys. Rev. 167 (1968) 1175-1179.

\bibitem{Novikov3} 
I.A. Dynnikov and S.P. Novikov, 
{\it Geometry of the triangle equation on two-manifolds},
axXiv:math-ph/0208041. 

\bibitem{flaschka} 
H. Flaschka, 
{\it On the Toda lattice. II Inverse scattering solutions},
Progr. Theoret. Phys. 51 (1974) 703-716.

\bibitem{Manakov} S.V.Manakov, {\it Complete integrability and stochastization 
in discrete dynamical systems}, Soviet Phys. JETP 40 (1975) 269-274.

\bibitem{Shabat} A.Shabat, in: Nonlinearity, Integrability and All That. Twenty 
Years After NEEDS'79, edited by M.Boiti, L.Martina, F.Pempinelli, B.Prinari and 
G.Soliani (Singapore: World Scientific, 2000) 331.

\bibitem{GLM} 
M. Grundland, D. Levi and L. Martina, 
{\it On a discrete version of the CP1 sigma model and surfaces immersed in R-3}
J. Phys. A-Math. Gen. 36 
%(16) 
(2003)  4599-4616. 

\end{thebibliography}
\end{document}